\documentclass[a4paper,12pt]{article}
\usepackage{amsthm}
\usepackage{amsfonts}
\usepackage{amssymb}
\usepackage{mathrsfs}
\usepackage[english]{babel}
\usepackage{float}
\usepackage[toc,page]{appendix}
\usepackage {graphicx}
\usepackage{epstopdf}
\usepackage{amsmath}
\usepackage{color}

\numberwithin{equation}{section}

\newtheorem{theorem}{Theorem}

\newtheorem{definition}[theorem]{Definition}

\begin{document}

\title{Exact recovery of non-uniform splines from the projection onto spaces of algebraic polynomials}
\date{July 2013, revised December 2013}
\author{Tamir Bendory (a), Shai Dekel (b), Arie Feuer (a) \footnote{(a) Department of Electrical Engineering, Technion, (b) GE Global Research and School of Mathematical sciences, Tel-Aviv University, Israel}}
\maketitle

\begin{abstract}
In this work we consider the problem of recovering non-uniform splines from their projection onto  spaces of algebraic polynomials. We show that under a certain Chebyshev-type separation condition on its knots, a spline whose inner-products with a polynomial basis and boundary conditions are known, can be recovered using Total Variation norm minimization. The proof of the uniqueness of the solution uses the method of `dual' interpolating polynomials and is based on \cite{SR}, where the theory was developed for trigonometric polynomials. We also show results for the multivariate case.

\end{abstract}

\section{Introduction}

In many applications, one obtains projections onto spaces of algebraic polynomials when using spectral methods to numerically solve partial differential equations (see e.g. \cite{spectral2}, \cite{Spectral1}). In the scenario where the solution contains discontinuities in higher order derivatives, even at a small number of separated locations, the convergence order of the spectral methods will be limited. An alternative to using higher degree polynomials in the numeric PDE method is to apply a `super-resolution' post-processing algorithm to a result in a lower degree space. 

In this work, we focus on the model problem of exact recovery of a non-uniform spline from a projection onto a space of algebraic polynomials. In the field of Compressed Sensing (CS), to make this type of problem tractable, one assumes `sparsity' of the unknown function (see for instance  \cite{Donoho_CS,uncertainty,universal_encoding,stable_recovery} and in particular \cite{CS_legendre} that deals with CS in the setting of `sparse' Legendre polynomials). However, sparsity is not sufficient for stable recovery. Therefore, following \cite{SR}, we place a Chebyshev-type separation condition on the location of the knots. Borrowing a term from the signal processing community, this  can be considered as a form of `finite rate of innovation' (but different than what  introduced in \cite{FRI}). The exact recovery of the splines is reduced into a problem of Total Variation norm minimization over Borel measures under given constraints. The proof of the uniqueness of the solution uses the method of `dual' interpolating polynomials \cite{SR}, \cite{beurling_extra}. 

We start by considering a complex measure $f$ of the form 
\begin{equation}
f=\sum_{m}c_m\delta_{x_m} ,\quad c_m\in\mathbb{C} , 
\label{eq:signal}
\end{equation}
where $X=\{x_m\}$ is the support of $f$ on $[-1,1]$, and $\delta_x$ is a Dirac measure.
Obviously, the measure $f$ can be interpreted as a linear functional, acting on any $g\in{C}[-1,1]$ as 
\begin{equation*}
\langle g,f\rangle=\sum_m c_m g(x_m) . 
\end{equation*}

Let $V_N$ be the space of univariate algebraic polynomial of degree $N$, and let 
 $\{P_k\}_{k=0}^{N}$ be any basis of $V_N$. For instance, $\{P_k\}_{k=0}^{N} $ may be selected as the standard basis $\{1,x,x^2,\dots,x^N\}$ or an orthonormal system such as Legendre polynomials. The goal is to recover $f$ from the set of its `inner-products' with $\{P_k\}_{k=0}^{N} $, i.e.
 \begin{equation}
 y_k = \langle f,P_k\rangle \,,\,  0\leq k\leq N .
 \end{equation}
 
The technique used in this paper is based on the approach of Cand\`{e}s and  Fernandez-Granda \cite{SR} who developed the theory for trigonometric polynomials. They proved that if the Fourier coefficients of a measure $f$ of type (\ref{eq:signal}) are known up to some given frequency and the Diracs are sufficiently separated (relative to the frequency), then $f$ is the unique minimizer of a Total-Variation (TV) norm over all complex measures that satisfy the constraints. Adapting the methodology of \cite{SR} to our setting requires to use the natural metric associated with algebraic polynomials over the interval $[-1,1]$, which is $\rho(x,y):=|\arccos(x)-\arccos(y)|$, $\forall x,y \in [-1,1]$ (see Chapter 8 in \cite{DL}). We then have

\begin{definition} \label{sep-def}
For $N \ge 128$, a set of points $X\subset \left[\cos\left(-\pi+\frac{2\pi}{N}\right),\cos\left(-\frac{2\pi}{N}\right)\right]$, is said to satisfy the minimal separation condition if any $x_i,x_j\in X$, obey $\rho(x_i,x_j)\geq \frac{4\pi}{N}$.
\label{def:distance}
\end{definition}

We note that the authors of \cite{beurling_extra}, also considered exact recovery in the setting of algebraic polynomials of degree $N$, but they imposed a minimal separation Euclidean distance of order $\mathcal{O}(N^{-0.4})$, whereas our results allow a Euclidean separation of order $\mathcal{O}(N^{-1})$ and even allow the order of $\mathcal{O}(N^{-3/2})$ near the endpoints.

\begin{definition}
Let $\mathcal{B}(A)$ be the Borel $\sigma$-Algebra on $A\subset \mathbb{R}^n$, where $A$ is compact and denote by $\mathcal{M}(A)$ the associated space of complex Borel measures. The total variation of a complex Borel measure $v\in \mathcal{M}(A)$ over a set $B\in\mathcal{B}(A)$ is defined by 
\begin{equation*}
\vert v\vert (B)=\sup\sum_k\vert v(B_k)\vert,
\end{equation*}
where the supremum is taken over all partitions of $B$ into a finite number of disjoint measurable subsets. The total variation $\vert v \vert$ is a non-negative measure on $\mathcal{B}(A)$, and the Total Variation (TV) norm of $v$ is defined as
\begin{equation*}
\|v\|_{TV}=\vert v\vert (A).
\end{equation*} 
\end{definition}
\noindent For a measure of the form of (\ref{eq:signal}), it is easy to see that
\begin{equation} 
\|f\|_{TV}=\sum_m\vert c_m\vert.
\end{equation}

\noindent Equipped with the definitions above, we are ready to state our first main result.
\begin{theorem}
Let $X:=\{x_m\}$ be the support of a complex measure $f$ of the form (\ref{eq:signal}). 
Let $\{P_k\}_{k=0}^N$ be any basis of $V_N$ for $N\geq 128$, and let $y_k = \langle f,P_k\rangle$ for all $0\leq k\leq N$. If the set X satisfies the separation condition of Definition \ref{sep-def}, then $f$ is the unique solution of
\begin{equation}
\min_{g\in\mathcal{{M}}([-1,1]) } \|g\|_{TV} \quad \mbox{subject to} \quad y_k = \langle g,P_k\rangle  \,,\, 0\leq k\leq N.
\label{eq:TV_min}
\end{equation}
\label{th:spikes}
\end{theorem}
The theorem states that if the support of $f$ is sufficiently separated, then it is the unique complex measure which is consistent with the measurements, and has minimal TV norm. We emphasize that this theorem can readily be extended to any finite interval. Also, observe that if the coefficients $\{c_m\}$ are known to be real and positive, then no separation condition is in fact needed and $f$ can be recovered uniquely by TV minimization (over all non-negative measures) as long as the number of Diracs $\le N/2$ \cite{beurling_extra}.

Theorem \ref{th:spikes} can be extended to higher dimensions. Here, we give a concrete example for the two-dimensional case. We consider a real bivariate measure of the form
\begin{equation}
f=\sum_{m}c_m\delta_{x_m}, 
\label{eq:2d_signal}
\end{equation}
 where $X:=\{x_m\}\subset (-1,1)^2$, and $c_m\in\mathbb{R}$ are coefficients. 
Before stating the theorem, we introduce the two-dimensional version of the separation condition as follows:
\begin{definition}
For $N \ge 512$, a set of points $X\subset \left[\cos\left(-\pi+\frac{2\pi}{N}\right),\cos\left(-\frac{2\pi}{N}\right)\right]^2$ satisfies the 2D-minimal separation condition if any $x_i,x_j\in X$ obey
\begin{equation*}
\max\left\{ {\rho(x_i(1),x_j(1)),\rho(x_i(2),x_j(2))}\right\} \geq \frac{4.76\pi}{N}.
\end{equation*}
\label{def:2d}
\end{definition}

\begin{theorem}
Let $X:=\{x_m\}$ be the support of a real bivariate measure $f$ of the form (\ref{eq:2d_signal}).
Let $V_N^2$ be the space of bivariate polynomials of degree N, $N\geq 512$, and let $\{P_k\}_{k=1}^{(N+1)^2}$ be any corresponding basis. If X satisfies the 2D-minimal separation condition, then $f$ is the unique real measure with minimal TV norm, that satisfies the constraints $y_k = \langle f,P_k\rangle$ for all $1\leq k\leq (N+1)^2$.
\label{th:2d}
\end{theorem}

%\subsection{Extension to Splines}

Our main application of Theorem \ref{th:spikes} is to the case of projections of non-uniform splines onto spaces of algebraic polynomials. A univariate spline of degree $r$ over the knot sequence $\{-1,x_1,\dots,x_M,1\}$, is a an ($r-1$) times continuously differentiable function of the form
\begin{equation}
f(x)=\mathbf{1}_{[-1,x_1)}(x)p_0(x) + \sum_{m=1}^{M-1} {\mathbf{1}_{[x_{m},x_{m+1})}(x)p_m(x)} + \mathbf{1}_{[x_M,1]}(x)p_M(x),
\label{eq:spline}
\end{equation}
where $p_m$, $m=0,\dots,M$, are polynomials of degree $r$ (with possibly complex coefficients), and 
\begin{equation*}
\mathbf{1}_{A}(x)= \begin{cases} 1 &  x\in A, \\
0 &  x\notin A. \end{cases}
\end{equation*}

\begin{theorem}
Assume that for a univariate (possibly complex) spline $f$ of degree $r$:
\begin{enumerate}
\item [(i)] The knots $X:=\{x_m\}$, $m=1,\dots,M$, satisfy the minimal separation condition of Definition \ref{sep-def} for $N\geq 128$,
\item [(ii)] The projection of $f$ onto $V_N$ is known,
\item [(iii)] The boundary conditions $\left\{f^{(j)}(-1),f^{(j)}(1)\right\}\,,\, j=0,\dots,r-1$, are given.
\end{enumerate}
 Then $f$ can be uniquely recovered through TV minimization.
%\begin{equation}
%\min_{g\in\mathcal{B}([-1,1]) } \|g\|_{TV} \quad \mbox{subject to} \quad y_k^r = \langle g,P_k\rangle  \,,\, 0\leq k\leq N
%\end{equation}
\label{th:main}
\end{theorem}

The structure of the paper is as follows. In Section 2 we review the `dual' problem of polynomial interpolation. In Section 3 we present the proofs of the main results. Finally, in Section 4 we show how the algebraic recovery problem can be easily recast into the trigonometric case where existing recovery algorithms via convex optimization  \cite{SR,SR2} or the Prony method  \cite{Fourier_samples} can be used.

\section{The Dual Interpolating Polynomials}

Following \cite{SR}, our results are based on the construction of `dual' interpolating polynomials. Here, we present this principle in a more general form. 

\begin{theorem}
Let $f=\sum_mc_m\delta_{x_m}$ where $X:=\{x_m\}\subseteq A$, and $A\subset \mathbb{R}^n$ is compact. Let $\Theta_D$ be a linear space of continuous functions of dimension $D+1$ in $A$. For any basis $\{\theta_k\}_{k=0}^D$, of $\Theta_D$, let $y_k = \langle f,\theta_k\rangle$ for all $0\leq k\leq D$. If for any set $\{ u_m \}$, $u_m\in\mathbb{C}$, with $\vert u_m\vert=1$, there exists $q\in \Theta_D$ such that
\begin{align}
q(x_m)&=u_m \,,\, \forall x_m\in X  ,\label{eq:dual1} \\
\vert q(x)\vert&<1 \,,\, \forall x\in A\backslash X, \label{eq:dual2}
\end{align}
then $f$ is the unique complex Borel measure satisfying 
\begin{equation}
\min_{g\in\mathcal{{M}}(A) }\|g\|_{TV} \quad \mbox{subject to} \quad y_k = \langle g,\theta_k\rangle  \,,\, 0\leq k\leq D.
\label{eq:TV_min_duality}
\end{equation}

\begin{proof}
The proof is a generalization of the proof in the Appendix of \cite{SR}. It is a slight variant of Lemma 1.1 in \cite{beurling_extra}, 
but we give it here for the sake of completeness. Let $g$ be a solution of (\ref{eq:TV_min_duality}), and define $g=f+h$. The difference measure $h$ can be decomposed relative to $\vert f\vert$ as 
\begin{equation*}
h=h_X+h_{X^C},
\end{equation*}
where $h_X$ is concentrated in $X$, and $h_{X^C}$ is concentrated in $X^C$ (the complementary of $X$).
Performing a polar decomposition of $h_X$ yields
\begin{equation*}
h_X=\vert h_X\vert e^{i\phi(x)},
\end{equation*}
where $\phi(x)$ is a real function on $A$ (see e.g. \cite{rudin}). By assumption, there exists $q\in \Theta_D$ obeying 
\begin{align}
q(x_m)&=e^{- i\phi(x_m)} \,,\, \forall x_m\in X ,\label{eq:q_xm}\\
\vert q(x)\vert&<1 \,,\, \forall x\in A\backslash X . \label{eq:q_x}
\end{align}
Also by assumption $\langle g,\theta_k\rangle = \langle f,\theta_k\rangle$, for $0\leq k\leq D$, and so
\begin{equation}
\langle q,h\rangle = 0.
\label{eq:in_pro}
\end{equation}
The decomposition of $h$, the polar decomposition of $h_X$, (\ref{eq:q_xm}) and (\ref{eq:in_pro}) imply
\begin{equation*}
0=\langle q,h_X\rangle+\langle q,h_{X^C}\rangle=\|h_X\|_{TV}+\langle q,h_{X^C}\rangle.
\end{equation*}
If $h_{X^C}=0$, then $\|h_X\|_{TV}=0$, and $h=0$. Alternatively, if $h_{X^C}\neq 0$ ,we conclude by property (\ref{eq:q_x}) that 
\begin{equation*}
\vert \langle q,h_{X^C}\rangle \vert  < \|h_{X^C}\|_{TV}.
\end{equation*}
Thus, 
\begin{equation}
\|h_{X^C}\|_{TV}>\|h_{X}\|_{TV}.
\label{eq:nsp}
\end{equation}
Observe that (\ref{eq:nsp}) is similar to the discrete Null Space Property, presented in \cite{CS_bestK}, used to guarantee the success of the $\ell_1$ minimization method (see also \cite{beurling_extra}).
As a result of (\ref{eq:nsp}), we get 
\begin{equation*}
\begin{split}
\|f\|_{TV}&\geq \|f+h\|_{TV}=\|f+h_X\|_{TV}+\|h_{X^C}\|_{TV} \\
&\geq \|f\|_{TV}-\|h_X\|_{TV}+\|h_{X^C}\|_{TV}>\|f\|_{TV} ,
\end{split}
\end{equation*}
which is a contradiction. 
Therefore, $h=0$ which implies that $f$ is the unique solution of (\ref{eq:TV_min_duality}). 
\end{proof}
\label{th:duality}
\end{theorem}

\section{Proofs of the main results}

Consider a set of locations ${T}:=\{t_m\}\subset[-\pi,\pi]$, that satisfy the separation condition $\vert t_i-t_j\vert\geq \frac{4\pi}{N}$,  $\forall t_i,t_j\in{T}$. Here we consider the periodic (warp-around) distance. To be clear, the distance between $t_1=-\pi+0.1$ and $t_2=\pi-0.1$ is $0.2$. Using the good localization properties of the Jackson Kernel, the authors of \cite{SR}, were able to construct for any set $\{ u_m \}$, $u_m\in\mathbb{C}$, with $\vert u_m\vert=1$, a trigonometric polynomial $Q$ satisfying
\begin{align}
Q(t_m)&=u_m \,,\, \forall t_m\in{T} ,\label{eq:tri_dual1}\\
\vert Q(t)\vert&<1 \,,\, \forall t\in [-\pi,\pi]\backslash {T} \label{eq:tri_dual2},
\end{align}

Observe that if a set $X=\{x_m\}\subset(-1,1)$, satisfies the separation condition of Definition \ref{def:distance}, then the set $\tilde{T}:=\{\tilde{t}_m\}=\{\arccos(x_m)\}\subset[-\pi,0]$, satisfies the separation condition of \cite{SR}, e.g. any $\tilde{t}_i,\tilde{t}_j\in \tilde{T}$ satisfy $\vert \tilde{t}_i-\tilde{t}_j\vert\geq \frac{4\pi}{N}$.

\subsection{Proof of Theorem \ref{th:spikes}}
\begin{proof}
According to Theorem \ref{th:duality}, a sufficient condition for $f$ to be the unique solution of (\ref{eq:TV_min}) is the existence of an algebraic polynomial $P$ of degree $N$, satisfying (\ref{eq:dual1}) and (\ref{eq:dual2})
for any set of interpolating values $\{ \tilde{u}_m \}$, $\vert \tilde{u}_m\vert=1$.
To prove the existence of such $P$, we consider the transformation 
\begin{equation*}
t=arccos(x),
\end{equation*}
mapping the interval  $[-1,1]$  to $[-\pi,0]$, and each location $x_m$ to $\tilde{t}_m=\arccos(x_m)$, $m=1,\dots, M$.
We then construct a new sequence $T:=\{{t}_m\}_{m=1}^{2M}\subset [-\pi,\pi]$, by performing a mirror reflection of the knots $\{\tilde{t}_m \}$, about the point 0,
\begin{equation}
t_m:=\begin{cases} 
\tilde{t}_m &\quad m=1,\dots,M, \\
-\tilde{t}_{2M-m+1} &\quad m=M+1,\dots,2M.
\end{cases}
\label{eq:t_tilde}
\end{equation}
Observe that since $X\subseteq \left[\cos\left(-\pi+\frac{2\pi}{N}\right),\cos\left(-\frac{2\pi}{N}\right)\right] $, the distance between ${t}_{2M}$ and ${t}_{1}$, and between ${t}_{M}$ and ${t}_{M+1}$ is at least $\frac{4\pi}{N}$.
Similarly, we use reflection to construct a sequence of interpolation values $\left\{{u_m}\right\}_{m=1}^{2M}$ as
\begin{equation}
u_m:=\begin{cases} 
\tilde{u}_m &\quad m=1,\dots,M, \\
\tilde{u}_{2M-m+1} &\quad m=M+1,\dots,2M. 
\end{cases}
\label{eq:u_tilde}
\end{equation} 

\noindent Since $X$ satisfies the separation condition of Definition \ref{sep-def}, the sequence $T$ satisfies the minimal separation condition of \cite{SR} and therefore there exists a trigonometric polynomial 
$\tilde{Q}$, satisfying (\ref{eq:tri_dual1}) and (\ref{eq:tri_dual2}).

We define 
\begin{equation}
Q(t):=\frac{\tilde{Q}(t)+\tilde{Q}(-t)}{2},
\label{eq:q_tilde}
\end{equation}
which is an even trigonometric polynomial, and thus has the form  $Q(t)=\sum_{k=0}^{N}\beta_k\cos(k{t})$. 
Due to the construction by reflection of $\{{t}_m\}$ and $\{{u}_m\}$ in (\ref{eq:t_tilde}) and (\ref{eq:u_tilde}), $Q$ also satisfies properties (\ref{eq:tri_dual1}) and (\ref{eq:tri_dual2}). 

We may now conclude, that if $X$ satisfies the separation condition of Definition \ref{sep-def}, then there exists an algebraic polynomial 
\begin{equation*}
P(x):=Q(\arccos(x))=\sum_{k=0}^{N}\beta_k\cos(k\arccos(x)) \, ,\,x\in[-1,1],
\end{equation*}
 satisfying (\ref{eq:dual1})  and (\ref{eq:dual2}). This completes the proof. 
\end{proof}

\begin{figure}[ht]
\begin{center}$
\begin{array}{cc}
\includegraphics[width=0.5\textwidth,height=0.4\textwidth]{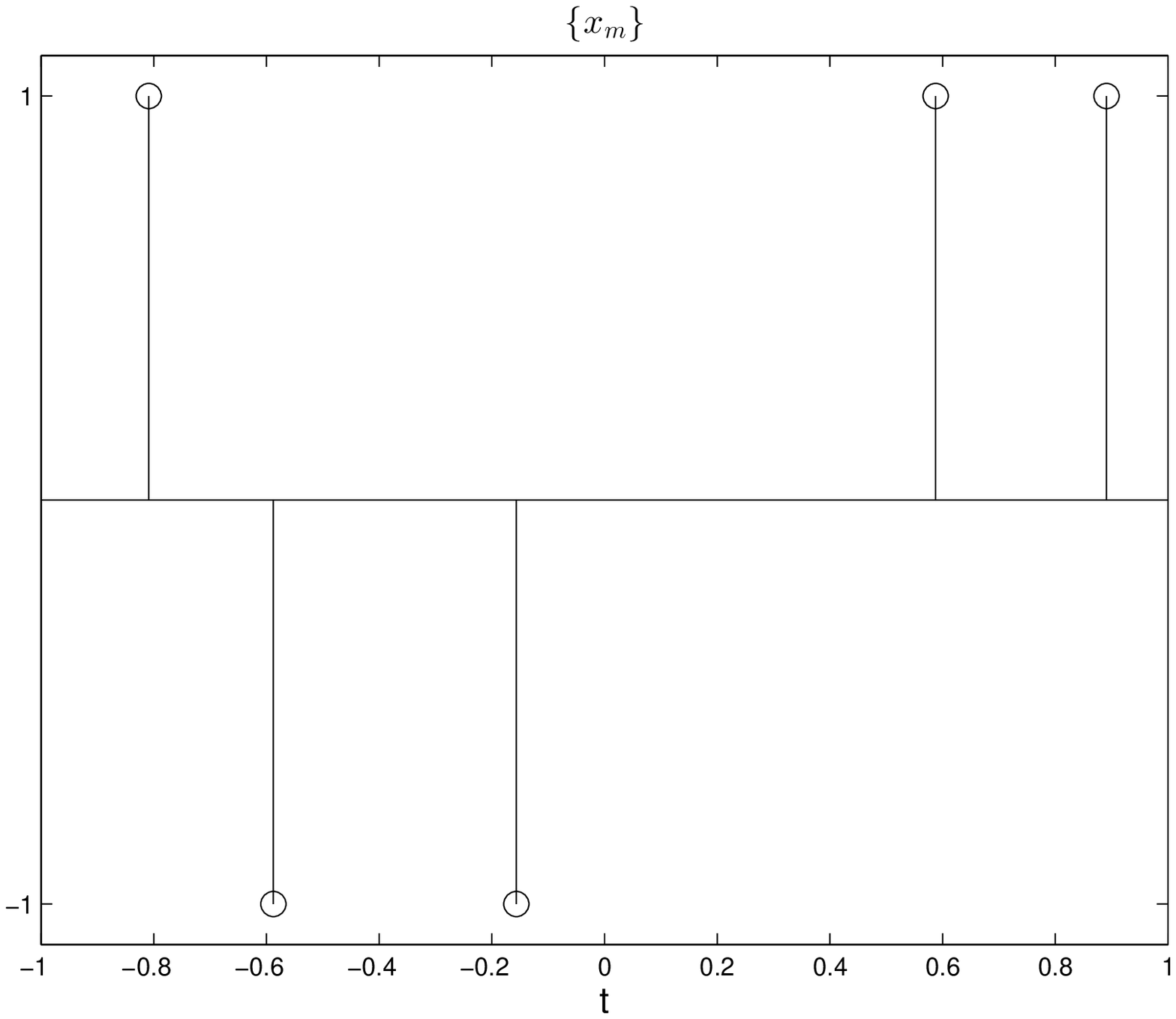} &
\includegraphics[width=0.5\textwidth,height=0.4\textwidth]{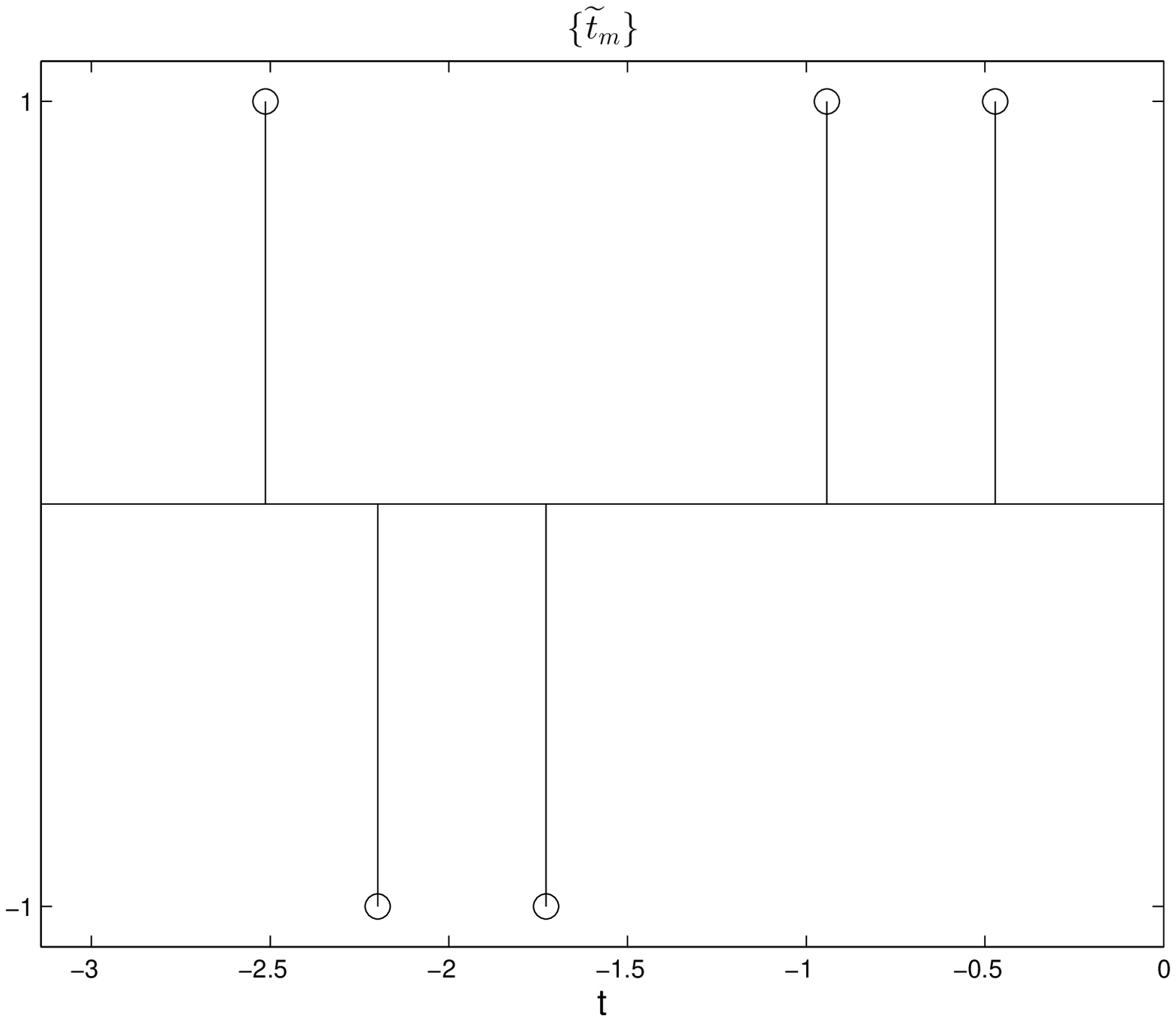} \\
\mbox{(a)} & \mbox{(b)} \\
\includegraphics[width=0.5\textwidth,height=0.4\textwidth]{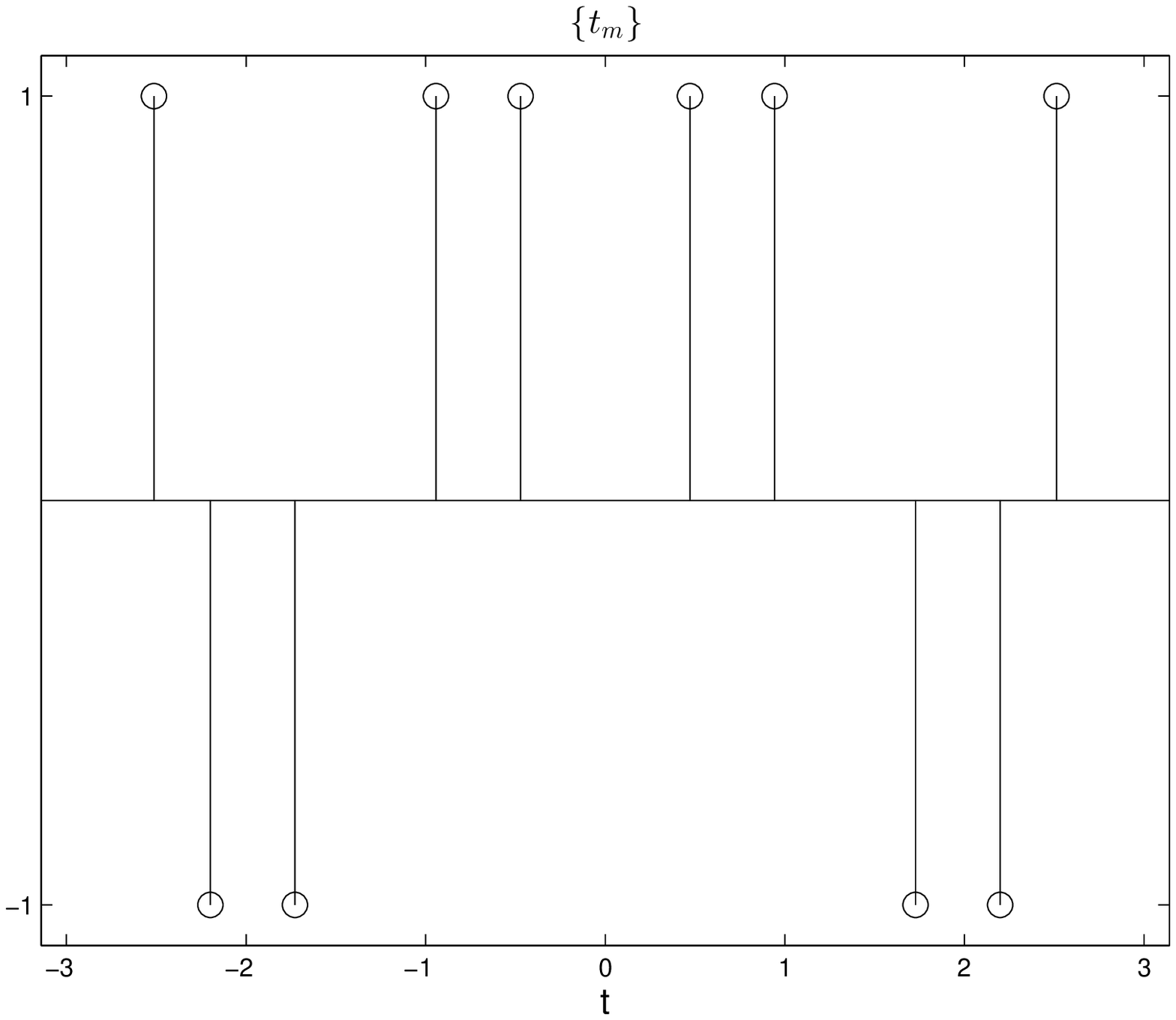} &
\includegraphics[width=0.5\textwidth,height=0.4\textwidth]{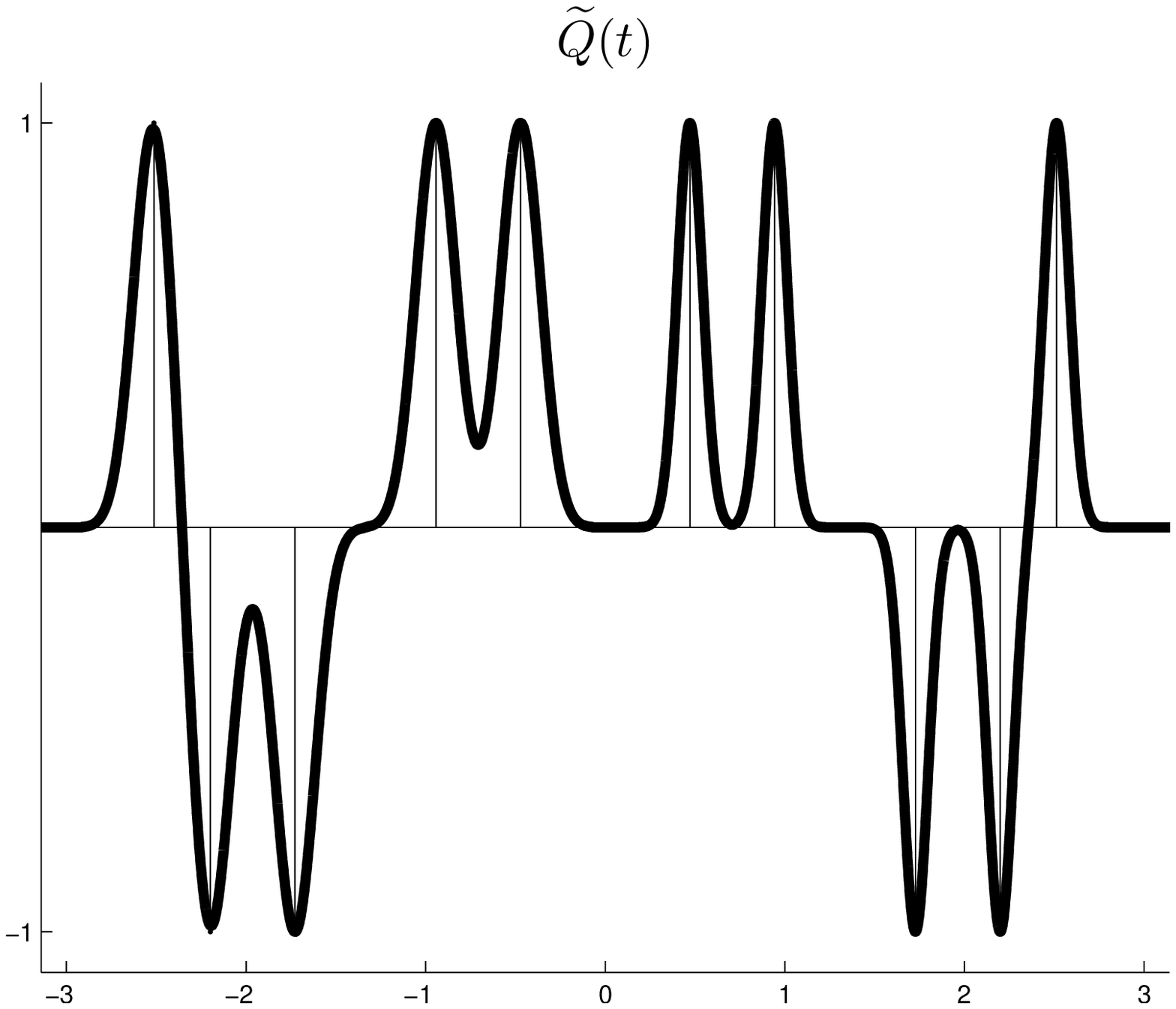} \\
\mbox{(c)} & \mbox{(d)} \\
\includegraphics[width=0.5\textwidth,height=0.4\textwidth]{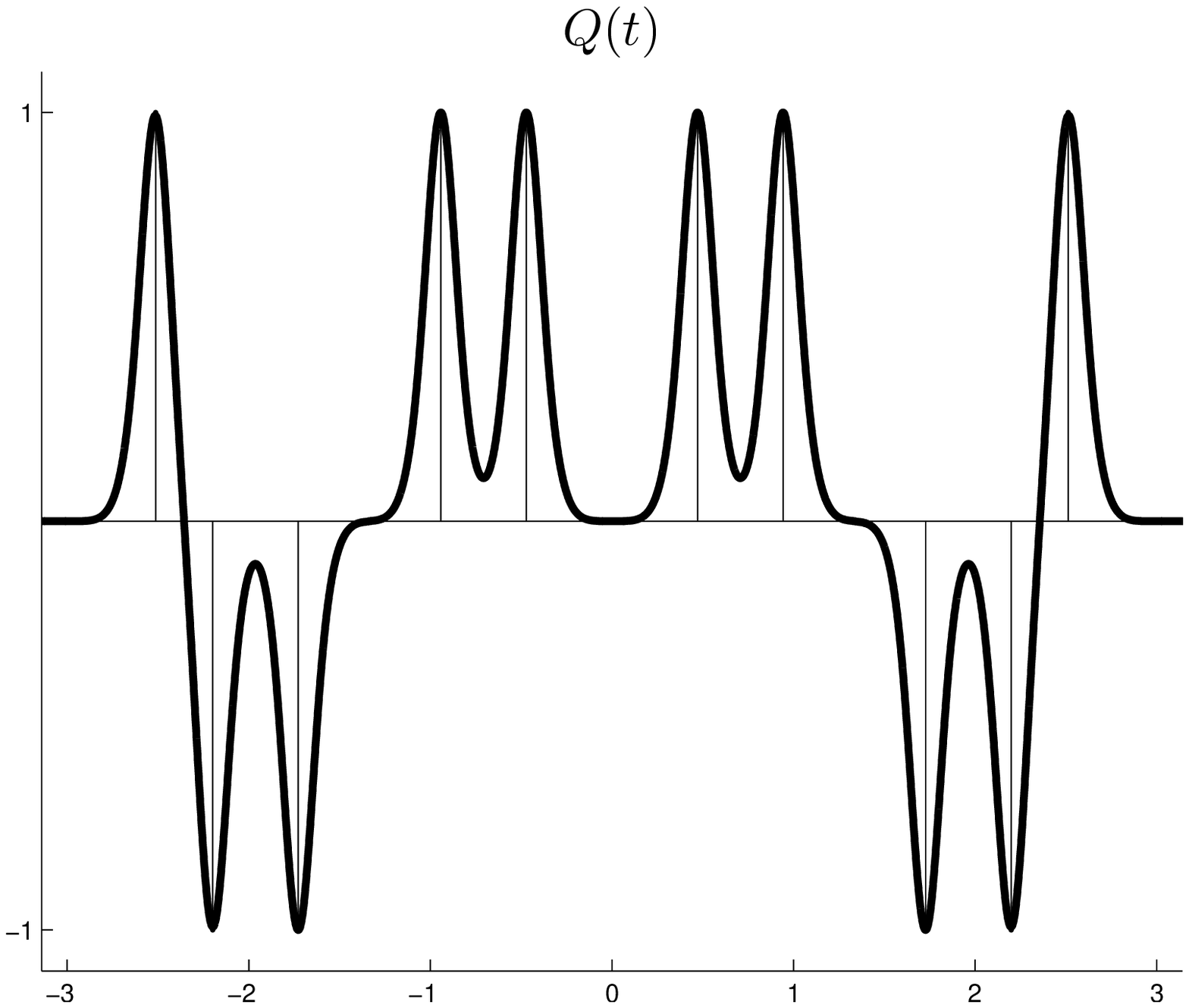} &
\includegraphics[width=0.5\textwidth,height=0.4\textwidth]{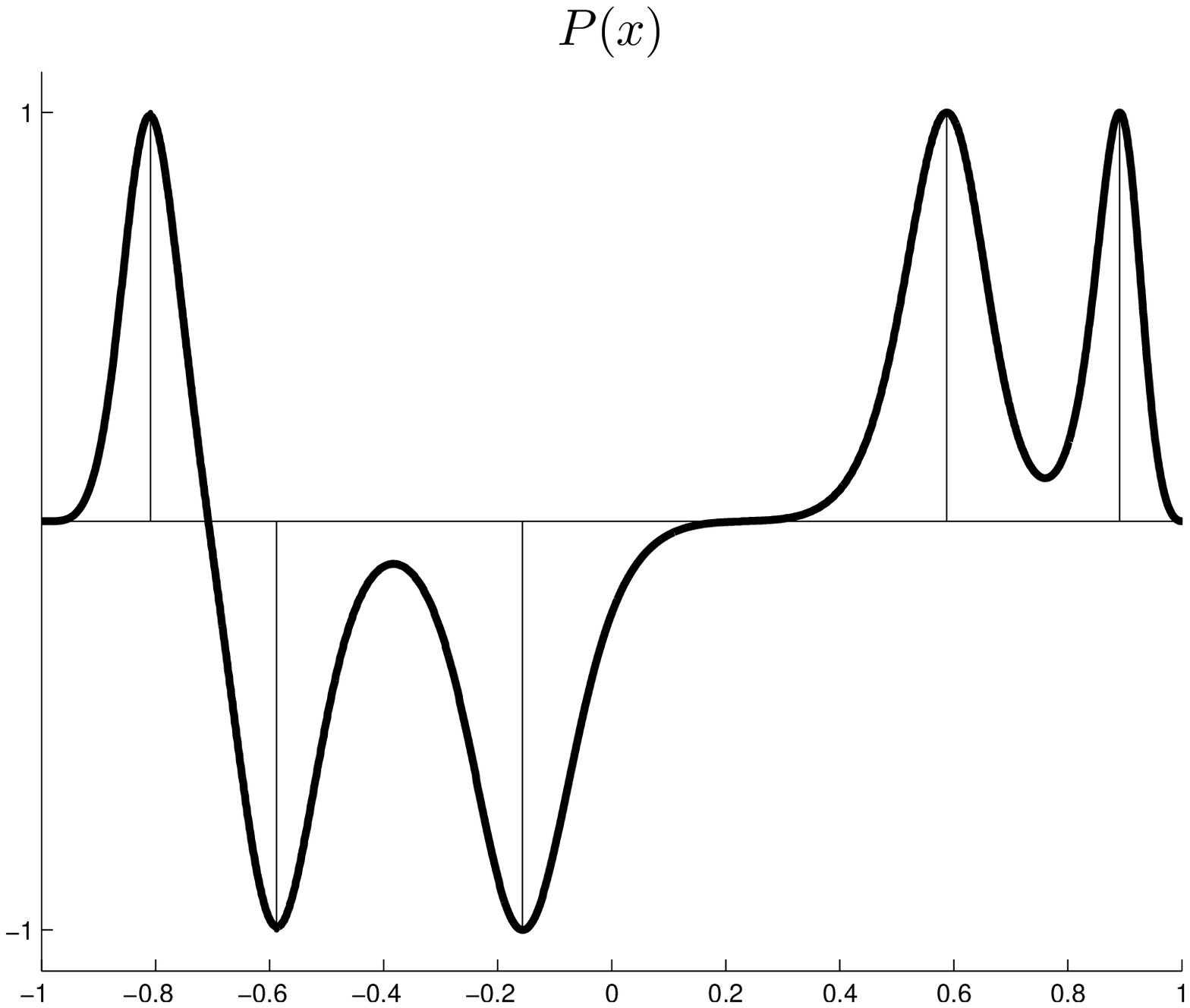} \\
\mbox{(e)} & \mbox{(f)} \\
\end{array}$
\end{center}
\caption{ (a) Arbitrary locations $\{{x}_m\}$ on $[-1,1]$ (b) $\{\tilde{t}_m\}=\{\arccos (x_m)\}$ on $[-\pi,0]$ (c) The extension $\{{t}_m\}$  on $[-\pi,\pi]$ according to (\ref{eq:t_tilde}) (d) the trigonometric polynomial $\tilde{Q}$, interpolating $\{{t}_m\}$ (e) the even trigonometric polynomial $Q$ (f) the algebraic polynomial $P(x)=Q(\arccos x)$. }
\label{fig:construction}
\end{figure}

The construction of the dual interpolating algebraic polynomial $P$ is demonstrated in Figure \ref{fig:construction}. 
For simplicity, we choose $\{\tilde{u}_m\}$ to be real valued.
Figure  \ref{fig:construction}(a) shows a separated set of locations $\{{x}_m\}\subset (-1,1)$ with associated interpolating values $\{\tilde{u}_m\}$. 
Figure \ref{fig:construction}(b) represents the set $\{\tilde{t}_m\}=\{\arccos(x_m)\}$.
Figure \ref{fig:construction}(c) shows the extended sequence $\{{t}_m\}\subset [-\pi,\pi]$ (see (\ref{eq:t_tilde}) and (\ref{eq:u_tilde})). In Figures  \ref{fig:construction}(d) and \ref{fig:construction}(e), one can see an example for a trigonometric polynomials $\tilde{Q}$ satisfying (\ref{eq:tri_dual1}) and (\ref{eq:tri_dual2}), and the even trigonometric polynomial $Q$, constructed by averaging (\ref{eq:q_tilde}). Ultimately, Figure \ref{fig:construction}(f) shows the algebraic polynomial $P(x)=Q(\arccos x)$.

In the above proof, we in fact used the classic approach of recasting an `approximation theoretical' problem in algebraic polynomials to the trigonometric case. Then, we significantly leveraged on the construction in \cite{SR} of interpolating trigonometric polynomials, under the separation assumption. In the realm of trigonometric polynomials, the good localization properties of Fej\'{e}r-type kernels provide a way to construct interpolating polynomials with the required properties. One uses translation invariance of these kernels and takes a superposition of the translates of the kernel and its derivative with centers at the prescribed knots in $[-\pi,\pi]$. Observe that one could in fact construct directly such a superposition of algebraic polynomials that are well-localized at prescribed knots in $[-1,1]$, but unlike the trigonometric case, there is no translation invariance and each such well-localized algebraic polynomial will depend on the location of the associated knot, where localization is with respect to the natural metric $\rho$.

The localization principal that is a key to the construction of interpolating polynomials can be extended under certain conditions to orthogonal polynomial systems on other compact manifolds. In \cite{recoverysphere} we demonstrate how this principle can also be applied in the case of Spherical Harmonics.

\subsection{Proof of Theorem \ref{th:2d}}
By Theorem \ref{th:duality}, the existence of a bivariate algebraic polynomial  $P$ satisfying (\ref{eq:dual1}) and (\ref{eq:dual2}), for any prescribed real interpolation values $\{\tilde{u}_m\}$, $\vert \tilde{u}_m\vert=1$, is a sufficient condition for the exact recovery of $f$ from the set of $(N+1)^2$ measurements $\{y_k\}$.   

The authors of \cite{SR} prove that if a set of locations $T:=\{t_m\}\subset[-\pi,\pi]^2$ satisfies
\begin{equation}
\max\left \{\left\vert t_i(1)-t_j(1) \right\vert,\left\vert t_i(2)-t_j(2) \right\vert\right\}\geq \frac{5.76\pi}{N}  \,,\, \forall t_i,t_j\in T \label{eq:t_2d},
\end{equation}
and $N\geq 512$, then there exists a bivariate trigonometric polynomial $Q(t)=\sum_{k_1,k_2=-N}^N a_{k_1,k_2}e^{ik_1t_1}e^{ik_2t_2}$ , obeying 
\begin{align}
Q(t_m)&=u_m \,,\, \forall t_m\in T ,\label{eq:2dtri_dual1}\\
\vert Q(t)\vert&<1 \,,\, \forall t\in [-\pi,\pi]^2\backslash T \label{eq:2dtri_dual2},
\end{align}
 for any set $\{u_m\}$, $u_m\in\mathbb{R}$ with $\vert u_m\vert =1$.

Let $X:=\{x_m\}\subset(-1,1)^2$, be the support of $f$, satisfying the separation condition of Definition \ref{def:2d}. We apply the transformation $(t_1,t_2)=(\arccos\left(x_1\right),\arccos\left((x_2\right))$, converting the locations $\{x_m\}\subset[-1,1]^2$ to $\tilde{T}:=\{\tilde{t}_m\}\subset[-\pi,0]^2$. We also associate each interpolation value $\tilde{u}_m$, with the transformed knot $\tilde{t}_m$. 
Similarly to (\ref{eq:t_tilde}) and (\ref{eq:u_tilde}), we symmetrize the locations $\tilde{T}$ and the interpolation values by reflecting the square $[-\pi,0]^2$ about the y-axis and then reflecting the rectangle $[-\pi,\pi]\times [-\pi,0]$ about the x-axis to obtain new symmetric locations $T:=\{{t}_m\}$ and prescribed values $\{u_m\}$, on the square $[-\pi,\pi]\times [-\pi,\pi]$. Observe that if the set $X\subset (-1,1)^2$ satisfies the 2D-minimal separation condition of Definition \ref{def:2d}, then the constructed symmetric set $T\subset [-\pi,\pi]^2$ satisfies (\ref{eq:t_2d}). Therefore, there exists a bivariate trigonometric polynomial $\tilde{Q}$ obeying (\ref{eq:2dtri_dual1}) and (\ref{eq:2dtri_dual2}).

Next, define a new trigonometric polynomial by
\begin{equation*}
Q(t_1,t_2):=\frac{1}{4}\left( \tilde{Q}(t_1,t_2)+ \tilde{Q}(t_1,-t_2)+\tilde{Q}(-t_1,t_2)+\tilde{Q}(-t_1,-t_2)\right).
\end{equation*}
Clearly, due to its construction through symmetry, $Q$ has the form of 
\begin{equation*}
Q(t_1,t_2)=\sum_{k_1,k_2=0}^Nb_{k_1,k_2}\cos\left(k_1t_1\right)\cos\left(k_2t_2\right).
\end{equation*}
Furthermore, $Q$ also satisfies (\ref{eq:2dtri_dual1}) and (\ref{eq:2dtri_dual2}) for the same locations $T=\{{t}_m\}$, and interpolation values $\{{u}_m\}$.

To conclude the proof, observe that if $X$ obeys the 2D separation condition of Definition \ref{def:2d}, then for any prescribed interpolation values $\{ \tilde{u}_m \}$, the bivariate algebraic polynomial $P(x)=Q(\arccos(x(1)),\arccos(x(2)))$ satisfies (\ref{eq:dual1}) and (\ref{eq:dual2}). Therefore, $f$ is the unique real measure with minimal TV norm, satisfying the data constraints.

\subsection{Proof of Theorem \ref{th:main}}

We assume that the knots of the spline, $X:=\{x_m\}$, $m=1,\dots,M$, satisfy the separation condition of Definition \ref{sep-def} for $N\geq 128$, and that the coefficients $y_k = \langle f,P_k\rangle$, $0\leq k\leq N$, and boundary conditions $\left\{f^{(j)}(-1),f^{(j)}(1)\right\}\,,\, j=0,\dots,r-1$, are known. 

Since $P'_k \in V_N$, $k=0 \dots N$, there exist coefficients $\{ \alpha_{k,n} \}$, $n=0 \dots N$, such that 
\begin{equation} \label{der-coef}
P'_k(x)=\sum_{n}\alpha_{k,n}P_n(x).
\end{equation}
 
\noindent Assume $f$ is a piecewise constant spline (i.e. $r=0$) 
\begin{equation}
f(x)=c_0\mathbf{1}_{[-1,x_1)}(x) + \sum_{m=1}^{M-1} {c_m\mathbf{1}_{[x_{m},x_{m+1})}(x)} + c_M\mathbf{1}_{[x_M,1]}(x).
\label{eq:piece-wise}
\end{equation}

\noindent The distributional derivative of  $f$ is 
\begin{equation}
f'(x)=\sum_{m=1}^M \delta_{x_m}(c_{m}-c_{m-1}).
\label{eq:piece-wise_der}
\end{equation}

\noindent Using (\ref{der-coef}) and the boundary conditions on $f$ allows to explicitly calculate
\begin{equation}
\begin{split}
\langle f',P_k\rangle&=\int_{-1}^1f'(x)P_k(x)dx=f(x)P_k(x)\Big\vert_{-1}^1-\langle f,P'_k\rangle  \\
&=f(1)P_k(1)-f(-1)P_k(-1)-\sum_n\alpha_{k,n}\langle f,P_n\rangle \\
&=f(1)P_k(1)-f(-1)P_k(-1)-\sum_n\alpha_{k,n}y_n.
\label{eq:int_parts}
\end{split}
\end{equation}

\noindent Since the knots of $f'$ satisfy the minimal separation condition, by Theorem \ref{th:spikes} it is the unique solution of
\begin{equation}
\min_{g\in\mathcal{{M}}([-1,1]) } \|g\|_{TV} \quad \mbox{subject to} \quad \langle f',P_k\rangle = \langle g,P_k\rangle  \,,\, 0\leq k\leq N .\label{eq:TV_spline0}
\end{equation}

\noindent Once $f'$ is determined uniquely from TV minimization, utilizing one of the boundary conditions (e.g. $f(-1)$) uniquely determines $f$. 

To extend this result to higher degrees, one uses induction on $r$. In general, suppose that $f$ is a spline of degree $r\ge 1$ of the form
\begin{equation}
f(x)=\mathbf{1}_{[-1,x_1)}(x)p_0(x) + \sum_{m=1}^{M-1} {\mathbf{1}_{[x_{m},x_{m+1})}(x)p_m(x)} + \mathbf{1}_{[x_M,1]}(x)p_M(x),\label{eq:spline_form}
\end{equation}
where $p_m$ are polynomials of degree $r$. Then, $f'$ satisfies the following:
\begin{enumerate}
  \item [(i)] It is a spline of degree $r-1$,
  \item [(ii)] Its knot sequence is identical to $f$, satisfying the separation condition,
  \item [(iii)] Its inner-products with the algebraic polynomial basis can be explicitly computed as in (\ref{eq:int_parts}). 
  \item [(iv)] Its boundary conditions at the endpoints $-1,1$ are obviously known from the boundary conditions of $f$.
\end{enumerate}

By the induction hypothesis, $f'$ can be uniquely determined from a TV minimization procedure. Consequently, we uniquely determine $f$ using the boundary conditions. Finally, observe that $f$, which is a spline of degree $r$, is in fact determined through the recursion process by obtaining the Dirac-train $f^{(r+1)}$, as the unique TV minimizer satisfying constraints determined from information known about $f$. Once $f^{(r+1)}$ is known, one `rolls' back from the recursion to recover $f$ using the boundary conditions.

\section{Recasting the recovery algorithm}

As demonstrated in the proof of Theorem \ref{th:main}, we can reduce the problem of recovering a spline case to the recovery of  
\[
f\left( x \right)=\sum\limits_m {c_m \delta_{x_m } \left( x \right)} , -1\le x\le 1,
\]
from the projection onto the polynomial space of degree $N$, given by
\[
y_k =\int\limits_{-1}^1 {f\left( x \right)x^kdx} \quad ,
\quad
k=0,\ldots ,N.
\]
One way to compute a concrete recovery of $f$ is to recast the problem to 
the trigonometric case where there are existing convex optimization 
algorithms \cite{SR,SR2}. We therefore proceed with the substitution $x=\cos t$, $-\pi \le t \le 0$,
\[
\begin{split}
 y_k &=-\int\limits_{-\pi }^0 {f\left( {\cos t} \right)\cos^kt\sin tdt} \\ 
 &=\frac{1}{2}\left( {\int\limits_{-\pi }^0 {f\left( {\cos t} \right)\cos 
^kt\sin \left( {-t} \right)dt} +\int\limits_0^\pi {f\left( {\cos \left( {-t} 
\right)} \right)\cos^k\left( {-t} \right)\sin \left( t \right)dt} } \right) 
\\ 
 &=\frac{1}{2}\int\limits_{-\pi }^\pi {f\left( {\cos t} \right)\sin \left( 
{\left| t \right|} \right)\cos^k\left( t \right)dt}. \\ 
 \end{split}
\]
Observe that by the assumptions of Definition \ref{sep-def}, the support of $f$ is sufficiently away 
from the end points $\pm 1$. Therefore, the symmetric function $f\left( {\cos t} 
\right)\sin \left( {\left| t \right|} \right)$, $-\pi \le t\le \pi $, is 
also a superposition of Diracs with
\begin{equation} \label{recast}
f\left( {\cos t} \right)\sin \left( {\left| t \right|} 
\right)=\sum\limits_{m=1}^{2M} {c_m \sin \left( {\left| {t_m } \right|} 
\right)\delta_{t_m } \left( t \right)} .
\end{equation}
Here, as in the proof of Theorem \ref{th:spikes}, one can define $\tilde{{t}}_m :=\mbox{arccos}\left( 
{x_m } \right)$, $-\pi <\tilde{{t}}_m <0$, and then $\left\{ {t_m } \right\}$ as  
the symmetric extension of $\left\{ {\tilde{{t}}_m } \right\}$. The reflected coefficients satisfy $c_m 
=c_{2M-m+1} $ , $m=M+1,\ldots ,2M$. Resolving the function (\ref{recast}) over $\left[ 
{-\pi ,\pi } \right]$, obviously also yields exact recovery of $f$.

\noindent {\bf Acknowledgment} We thank the referees for their valuable comments that have significantly improved this paper.

\end{document}